\documentclass[epj]{svjour}
\usepackage{latexsym}
\usepackage{graphics}
\usepackage{epsfig}
\begin{document}
\title{Precision measurement of vector and tensor analyzing powers in elastic deuteron-proton scattering}
\author{H.~Mardanpour\inst{1,}\thanks{{email: mardanpur@kvi.nl}}, H.R.~Amir-Ahmadi\inst{1}, A. Deltuva\inst{2},
 K.~Itoh\inst{3}, N.~Kalantar-Nayestanaki\inst{1}, T.~Kawabata\inst{4}, H.~Kuboki\inst{5}, Y.~Maeda\inst{4},
 J.G.~Messchendorp\inst{1,}\thanks{{email: messchendorp@kvi.nl}}, S.~Sakaguchi\inst{4}, H.~Sakai\inst{4,5},
 N.~Sakamoto\inst{4}, Y.~Sasamoto\inst{4}, M.~Sasano\inst{5}, K.~Sekiguchi\inst{6}, K.~Suda\inst{4}, Y.~Takahashi\inst{5},
 T.~Uesaka\inst{4}, H. Wita\l{a}\inst{7}, and K.~Yako\inst{5}
}
\institute{
 Kernfysisch Versneller Instituut (KVI), University of Groningen, Zernikelaan 25, 9747 AA, Groningen, The Netherlands\and
 Centro de Fisica Nuclear da Universidade de Lisboa, P-1649-003 Lisboa, Portugal\and
 Department of Physics, Saitama University, Saitama 338-8570, Japan\and
 Center for Nuclear Study, University of Tokyo, Tokyo, Japan\and
 Department of Physics, University of Tokyo, Tokyo, Japan\and
 Nishina Center, RIKEN, Tokyo, Japan\and
 M. Smoluchowski Institute of Physics, Jagiellonian University, Krak\'ow, Poland}
\date{Received: date / Revised version: date}

\abstract{
 High precision vector and tensor analyzing powers of elastic deuteron-proton ($\vec{d}+p$) scattering have been measured at
 intermediate energies to investigate effects of three-nucleon forces. Angular distributions  in the range of 
 70$^{\circ}$-120$^{\circ}$ in the center-of mass frame for incident-deuteron energies of $E_{d}^{lab} = 130$ and $180$ MeV
 were obtained using the RIKEN facility. The beam polarization was unambiguously determined by measuring the
 $^{12}{\rm C}(\vec{d},\alpha)^{10}{\rm B}(2^{+})$ reaction at 0$^{\circ}$. Results of the measurements are compared with
 state-of-the-art three-nucleon calculations. The present modeling of nucleon-nucleon forces and its extension to the three-nucleon
 system is not sufficient to describe the high precision data consistently and requires, therefore, further investigation.
} 
\titlerunning{Measurement of the dp Analyzing Powers at 130 and 180 MeV.}
\authorrunning{H. Mardanpour}
\maketitle
\section{Introduction}
 The properties of nuclei and the dynamics of nuclear reactions are determined predominantly by the pair-wise nucleon-nucleon (NN)
 interaction~{\cite{carlson}}. The longest-range two-nucleon force (2NF) is due to the exchange of a pion, an idea that goes back 
to the work of Yukawa in 1935~{\cite{Yukawa,stoks}}. At present, existing 2NF models provide an excellent description of the high-quality 
database of proton-proton and neutron-proton scattering and of the properties of the deuteron. For the simplest three-nucleon system, the  
triton, an exact solution of the three-nucleon Faddeev equations employing 2NFs clearly underestimates the experimental binding 
energy~{\cite{nogga}}, and therefore shows that 2NFs are not sufficient to describe the three-nucleon system accurately. For heavier 
systems, the deviations between the calculated and the measured binding energies become even larger~{\cite{pieper}}. Deficiencies of theoretical 
predictions based on pair-wise nucleon-nucleon potentials have been observed in three-nucleon scattering observables as well. For 
example, rigorous Faddeev calculations~{\cite{gloeckle}} solely based on modern NN interactions fail to describe high-precision 
differential cross sections in proton-deuteron elastic scattering at intermediate energies obtained at KVI~{\cite{ermisch03,ermisch05}} 
and at RIKEN~{\cite{kimiko}}. A large part of the previously described discrepancies can be resolved by introducing additional 
three-nucleon forces (3NF)~{\cite{witala98}}. 

Most of the present-day 3NFs are based on a refined version of the Fujita-Miyazawa force~{\cite{fuji}} in which a 2$\pi$-exchange 
mechanism is incorporated by an intermediate $\Delta$ excitation of one of the nucleons. Later, more refined ingredients have been 
added as in Urbana IX~{\cite{urbana}} and Tucson-Melbourne (TM')~{\cite{tuscon}} allowing for additional processes contributing 
to the rescattering of the mesons on an intermediate excited nucleon. A different approach is provided by the Hannover theory group,
 where the $\Delta$-isobar is treated on the same basis as the nucleon,
 resulting in a coupled-channel potential CD-Bonn+$\Delta$~{\cite{hannover}} with pair-wise nucleon-nucleon and nucleon-$\Delta$ 
interactions mediated through the exchange of $\pi$, $\rho$, $\omega$, and $\sigma$ mesons. Within this self consistent framework, the 
$\Delta$-isobar excitation mediates an effective 3NF with prominent Fujita-Miyazawa and Illinois ring-type contributions~{\cite{pieper}}.

A measurement of differential scattering cross sections in three-nucleon systems is one of the tools used to study the general nature of 
the 3NF. More detailed information can be obtained by measuring other observables such as analyzing powers. For instance, the 
spin-dependent part of the 3NF can be studied specifically by measuring the vector and tensor analyzing powers using polarized proton 
and deuteron beams. Precision measurements of the vector analyzing power of the proton in elastic proton-deuteron scattering have 
been performed at various beam energies ranging from 90 to 250 MeV~{\cite{ermisch05,bieber,ermisch01,hatana}}. Data on vector and 
tensor analyzing powers in $\vec d+p$ scattering are, however, scarce. Vector and tensor analyzing powers at various energies between 
75-187~{\cite{witala}}, 191, and 395~MeV~{\cite{garcon}} have been obtained in the past. Most of these data do not have sufficient
 precision and, therefore, not enough sensitivity to study 3NF effects. The only precision data covering a large angular range were
 obtained using a polarized deuteron beam with energies of 140, 200 and 270~MeV~{\cite{kimiko,cadman}}. For a systematic study of the
 3NF, a more extensive database with higher precision data is essential.

This paper describes a precision measurement of vector and tensor analyzing powers in $\vec d+p$ elastic scattering. The experiment 
was conducted at the RIKEN accelerator research facility using a polarized deuteron beam with energies of 130 and 180~MeV. The paper 
continues with a description of the experimental facility and the techniques used to measure the spin observables. The results will 
be compared and interpreted using modern Faddeev calculations as described earlier. The paper ends with a summary and conclusions.

\section{Experimental procedure and results}
Vector and tensor analyzing powers in $\vec d+p$ elastic scattering at $E_{d}^{lab}$ = 130 and 180 MeV were measured at the RIKEN accelerator 
research facility. The experiment was composed of two setups. In the first setup, the polarization of the deuteron beam was determined 
by measuring the $^{12}{\rm C}(\vec d,\alpha)^{10}{\rm B}(2^{+})$ reaction at 0$^{\circ}$ using the SMART spectrograph. In the 
second setup, asymmetries in elastic $\vec d+p$ scattering were measured simultaneously with the SMART measurements using the D-room 
in-beam polarimeter. The vector and tensor analyzing powers in $\vec d+p$ elastic scattering were subsequently obtained by combining 
the results from the first and second setups. A detailed description of the experimental method can be found in Ref.~{\cite{suda}}. 
In this paper, we only provide a brief summary of the procedure.

The vector and tensor polarized deuteron beams were provided by the atomic beam polarized ion source~{\cite{atom}}. In this 
experiment, the polarization states of the deuteron beam were switched between off-mode ($p_{Z},p_{ZZ}$) = (0,0), up-mode 
$(p_Z,p_{ZZ})$ = (1/3,+1), and down-mode $(p_Z,p_{ZZ})$ = (1/3,$-$1). The values between the brackets correspond to the maximum theoretical 
values of the vector ($p_{Z}$) and tensor ($p_{ZZ}$) polarizations. The three different polarization states were switched every 5 s 
by changing the RF transition fields  of the polarized ion source. The experimental polarization values were typically 60-80$\%$ of 
the maximum possible theoretical value. The beam intensity used during the experiment was typically 30~nA.

For the unambiguous determination of the beam polarization, we have exploited the relation between beam polarization and the 
$^{12}{\rm C}(\vec d,\alpha)^{10}{\rm B}(2^+)$ reaction cross section at 0$^\circ$. Following the Madison convention~{\cite{ohlson}}, 
the cross section of a scattered particle, $\sigma$, for a spin-1 beam impinging on an unpolarized target and with an angle
 $\beta$ between the spin direction and the beam direction ($\hat{z}$) and an azimuthal angle $\phi$ defined to be
$0^\circ$ for scattering to the left of the beam in the horizontal direction ($\hat{x}$) and 90$^\circ$ for scattering downward
 ($-\hat{y}$), can be written as:
\begin{eqnarray}
\hspace{-20mm} \sigma& = &\sigma_{0}\lbrack 1+\sqrt{3}iT_{11}(\theta)p_{Z}\cos\phi\sin\beta\nonumber\\
 &&+\frac{T_{20}(\theta)p_{ZZ}}{\sqrt{8}}(3\cos^{2}\beta-1)\nonumber\\
&&+\sqrt{3}T_{21}p_{ZZ}\cos\beta\sin\beta\sin\phi\nonumber\\
&&-\frac{\sqrt{3}}{2}T_{22}(\theta)p_{ZZ}\cos{2\phi}\sin^{2}\beta\rbrack,
\label{eq2}
\end{eqnarray}
where $iT_{11},T_{20},T_{22}$, and $T_{21}$ are the vector and various tensor analyzing powers, respectively, and $\theta$ is the polar scattering angle.
 The vector and tensor polarization of the beam are denoted as $p_{Z}$ and $p_{ZZ}$, 
respectively. The reaction cross section  for an unpolarized beam is given by $\sigma_0$.

 For scattering at $\theta_{c.m.} = 0^{\circ}$, the cross section relation can be simplified, since  
$iT_{11}(0^{\circ}) = T_{22}(0^{\circ}) = T_{21}(0^{\circ}) = 0$. In this case, Eq.~(\ref{eq2}) becomes, 
\begin{eqnarray}
\frac{\sigma}{\sigma_{0}} = 1+\frac{T_{20}(\theta)p_{ZZ}}{\sqrt{8}}(3\cos^{2}\beta-1).
\label{eq1}
\end{eqnarray}

Equation~(\ref{eq1}) allows us to determine the tensor polarization of the beam by measuring the ratio between the polarized and 
unpolarized cross sections at 0$^\circ$. As a pre-requisite, the angle $\beta$ should be known during the experiment. The angle, $\beta$, 
 of the beam polarization can be manipulated at the ion source. The resulting polarization is preserved by the single turn extraction of the beam. 
 In addition, the tensor analyzing power, $T_{20}$, should be known exactly and preferably at its maximum value. The latter is achieved
 by using the reaction $^{12}{\rm C}(\vec d,\alpha)^{10}{\rm B}(2^+)$ as a measure for the beam polarization~{\cite{okamura,kuehner}}.
 For symmetry reasons, this reaction has a maximum analyzing power of $T_{20} = 1/\sqrt{2}$ at $\theta_{c.m.} = 0^{\circ}$. In addition, 
the $2^+$ excited state can be identified uniquely by measuring the scattered $\alpha$ particle using the SMART spectrograph. In the 
ratio $\sigma/\sigma_{0}$ in Eq.~(\ref{eq1}), detector-related parameters, such as solid angle and efficiencies, cancel out since the 
same detector is used for the different polarization states.

\begin{figure}[!h]
\centering
\resizebox{0.45\textwidth}{!}{\includegraphics[angle = 0,width = .5\textwidth]{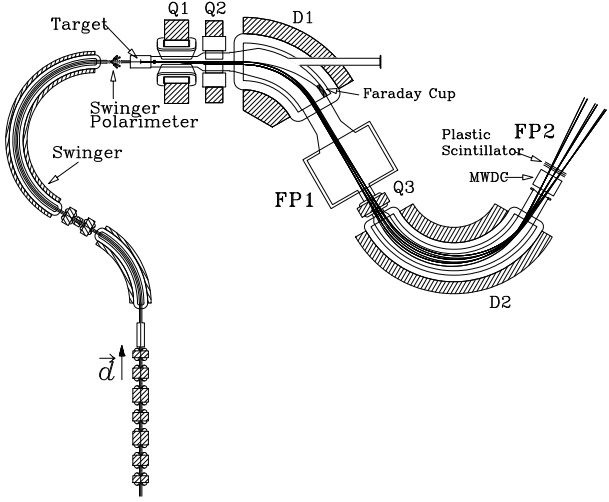}}
\caption{A schematic top view of the SMART spectrograph which was used to measure the 
$^{12}{\rm C}(\vec d,\alpha)^{10}{\rm B}(2^+)$ reaction at 0$^\circ$.}
\label{smart}
\end{figure}

The SMART magnetic spectrograph~{\cite{smart}} consists of a beam swinger and a cascade magnetic analyzer with two focal planes.
 The three quadruples and two dipole magnets were set in a QQDQD configuration. The scattered $\alpha$ particles were momentum analyzed 
by the magnetic spectrograph and detected by a multiwire drift chamber and three plastic scintillators placed at the second focal plane. 
A schematic top view of the SMART spectrograph is given in Fig. \ref{smart}. In this experiment the scattering of the polarized deuteron 
beam from a $^{12}$C target with a thickness of $20~\rm mg/cm^2$ was studied with the SMART spectrograph.

\begin{figure}[!h]
\centering
\resizebox{0.45\textwidth}{!}{\includegraphics[angle = 0,width = .4\textwidth]{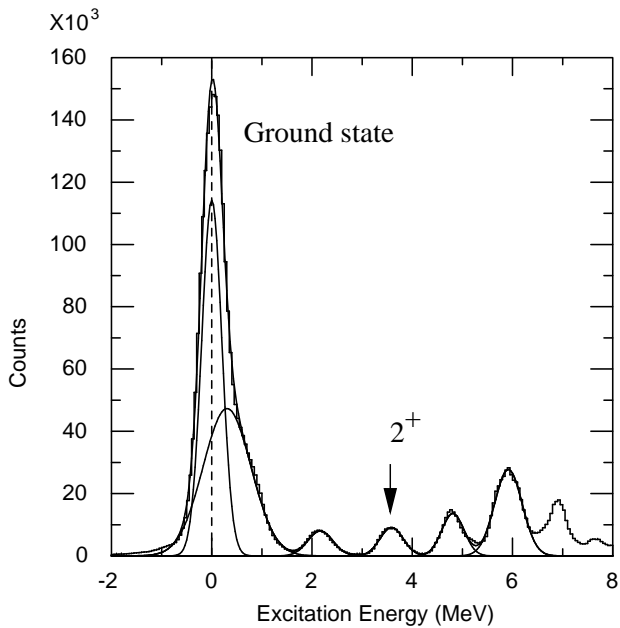}}
\caption{The excited states of $^{10}$B, obtained by a momentum analysis of the $\alpha$ particles in the reaction 
$^{12}{\rm C}(\vec d,\alpha)^{10}{\rm B}(2^+)$. The events stemming from the 2$^+$ state have been used to determine the beam 
polarization as described in the text.}
\label{excite}
\end{figure}

\begin{figure}[!h]
\centering
\resizebox{0.55\textwidth}{!}{\includegraphics[angle = 0,width = .8\textwidth]{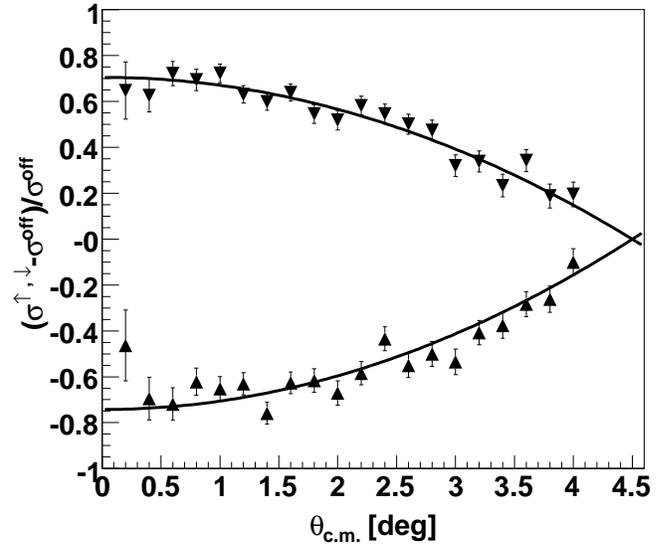}}
\caption{Polarization asymmetry of the $^{12}$C($\vec d,\alpha)^{10}$B(2$^+$) reaction 
as a function of the scattering angle in the center-of-mass frame. The two data sets were obtained from two polarization states 
of ($p_Z,p_{ZZ}$) = (1/3,+1) and ($p_Z,p_{ZZ}$) = (1/3,$-$1) normalized to the off polarization state  $(p_Z,p_{ZZ}) = (0,0)$. 
The asymmetry at $\theta_{c.m.} = 0^{\circ}$ is determined by an extrapolation of asymmetries measured up to 4$^{\circ}$ via a 
second order polynomial fit.}
\label{polAsym}
\end{figure}

Figure~\ref{excite} shows the excitation spectrum of $^{10}$B obtained by a momentum analysis of the scattered $\alpha$ particles. 
A clear signal from the 2$^+$ state can be identified. These events are used for the extraction of the tensor polarization of the 
deuteron beam according to Eq.~(\ref{eq1}).
The polarization asymmetries, $(\sigma^\uparrow - \sigma^{off})/\sigma^{off}$ and $(\sigma^\downarrow -\sigma^{off})/\sigma^{off}$, 
of the $(d,\alpha)$ reaction were obtained. The cross sections, $\sigma^\uparrow$, $\sigma^\downarrow$, and $\sigma^{off}$, 
correspond to a measurement with a beam polarization of $(p_Z,p_{ZZ})$ = (1/3,+1), $(p_Z,p_{ZZ})$ = (1/3,$-$1), and 
($p_{Z},p_{ZZ}$) = (0,0), respectively. Note that we used $\sigma^{off}$ instead of $\sigma_0$. In an ideal case, $\sigma^{off}$ 
corresponds to the cross section for a theoretically unpolarized beam. In our data analysis, however, a correction due to a 
small non-zero polarization in the $(p_Z,p_{ZZ})$ = (0,0) mode was taken into account. A typical angular distribution of the 
asymmetries of the above-mentioned reaction for two different polarization states is shown in Fig.~\ref{polAsym}. To obtain a 
precise asymmetry at 0$^\circ$, a second-order polynomial was fitted to the observed angular distribution as shown by 
the solid lines. Fluctuations in the beam polarization were monitored by measuring these asymmetries in 20 time slices of 60 
minutes each. These fluctuations were found to be small.

The D-room polarimeter at RIKEN was used to measure polarization asymmetries in the elastic $\vec d+p$ reaction. Combined with 
a simultaneous measurement of the beam polarization using the SMART spectrometer, vector and tensor analyzing powers in elastic 
$\vec d+p$ scattering are obtained. The D-room polarimeter consists of 4$\times$7 plastic scintillators placed to the right
 ($\phi$ = 0$^\circ$), left ($\phi$ = 180$^\circ$), up ($\phi$ = 90$^\circ$), and down ($\phi$ = 270$^\circ$) of a solid CH$_2$ target.
 The elastic $\vec d+p$ reaction was identified nearly background-free by a left-right or up-down coincidence requirement. For this,
 four scintillators were placed at a polar scattering angle of 25$^\circ$ covering a large part of deuteron phase space. These
 detectors were required to be in coincidence with one of the remaining smaller proton detectors placed on the other side of the
 beam at polar scattering angles between 30 and 55 degrees. 
\begin{figure}[!h]
\centering
\resizebox{0.45\textwidth}{!}{\includegraphics[angle = 0,width = .4\textwidth]{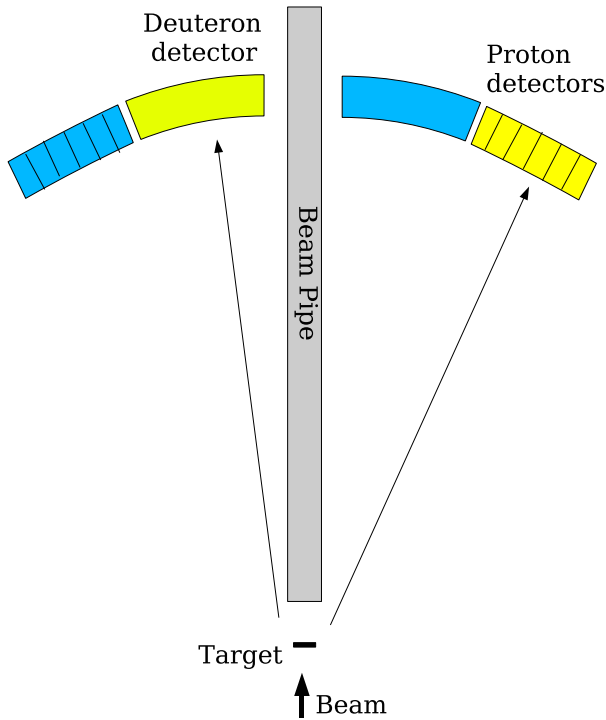}}
\caption{A sketch of the experimental setup used to measure the elastic deuteron-proton scattering process. Six small proton
 detectors are placed in coincidence with one deuteron detector placed on the other side of the beam. The coincidence setup in
 light gray (yellow) and in dark gray (blue) correspond to an azimuthal plane of $\phi$ = 0$^\circ$ (left) and $\phi$ = 180$^\circ$
 (right). The coincidence planes at $\phi$ = 90$^\circ$ and 270$^\circ$ are not shown.}
\label{droom}
\end{figure}
Figure~\ref{droom} illustrates schematically the coincidence setup for the left and right side on the horizontal plane. The vertical plane
 is not shown in this picture.

\begin{figure}[!h]
\centering
\resizebox{0.45\textwidth}{!}{\includegraphics[angle = 90,width = .4\textwidth]{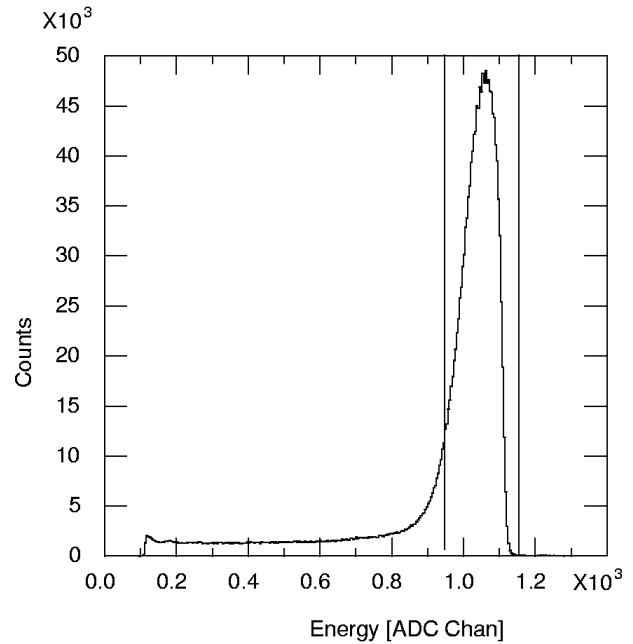}}
\caption{The response of one of the proton detectors placed at 30$^{\circ}$ and for a beam energy of 180~MeV. A clear signal
 stemming from the elastic deuteron-proton reaction can be observed with negligible background. The selected events used for the 
asymmetry measurement lie in the gate shown in the figure.}
\label{droomP1}
\end{figure}

 Figure~\ref{droomP1} shows the spectrum corresponding to the deposited energy of one of the proton detectors in coincidence
 with the opposite deuteron detector. This proton detector was placed at an angle of 30$^{\circ}$, and it had a thickness of 10 mm.
 All impinging protons punch through the detector and deposit an energy of about 15 MeV. The small tail on the left of the peak
 stems predominantly from hadronic interactions with the scintillator material. Furthermore, note that the pile-up background is
 negligible, which is indicated by the lack of events on the right side of the main peak. For the asymmetry determination in
 $\vec d+p$ scattering, events were selected within the gate shown in Fig.~\ref{droomP1}. 

The vector and tensor analyzing powers can be obtained by making use of the $\phi$ dependence according to Eq.~(\ref{eq2}).
The analyzing powers have been obtained by a global $\chi^2$-minimization fit of all data from the SMART and the D-room polarimeter,
 using Eq.~(\ref{eq2}). The free parameters in this fit were the analyzing powers, the tensor polarizations $p_{ZZ}^\uparrow$,
 $p_{ZZ}^\downarrow$, $p_{ZZ}^{off}$, and the cross section for an unpolarized beam, $\sigma_0$. Note that as a consequence of 
the operation mode of the atomic polarized-beam source, the vector polarization is 1/3 of the obtained tensor polarization, 
independent of the efficiencies of the transition units. The combined $\chi^2$ fitting procedure with 3 degrees of freedom 
was used to extract the experimental parameters and monitor them over time to estimate the systematic uncertainty.

With the geometry shown in Fig.~\ref{droom}, the vector and tensor analyzing powers in elastic $\vec d+p$ scattering were measured
 for $\theta_{c.m.} = 70^{\circ}-120^{\circ}$  at $E_{d} = 130$ MeV and $\theta_{c.m.} = 76^{\circ}-120^{\circ}$  at $E_{d} = 180$ MeV.
 The results for $E_d^{lab}$ = 130~MeV are shown in Fig.~\ref{130results} and listed in Tab.~\ref{130table}. Similarly, the results
 for $E_d^{lab}$ = 180~MeV can be found in Fig.~\ref{180results} and in Tab.~\ref{180table}. The error bars in Figs.~\ref{130results}
 and~\ref{180results} indicate only statistical uncertainties together with a small point-to-point (PTP) uncertainty. The small PTP
 error is added to account for small, non statistical, fluctuations of the data points as judged by comparison to a fit using a 4th
 order polynomial. These PTP uncertainties are added quadratically to the statistical uncertainties. 

 The dominant systematic uncertainty stems from the knowledge of the polar, $\beta$, and azimuthal, $\alpha$, angles of the
 polarization direction with respect to the beam direction. These angles were determined in a separate measurement which was
 conducted immediately after the $(\vec d,\alpha)$ and $\vec d+p$ measurements. In this measurement, the polarization axis precesses in
 the swinger dipole magnet of the SMART spectrograph which was rotated by $90^{\circ}$ in a vertical plane. The precession angle
 is measured before and after the spin precession by using two polarimeters. For a beam energy of 130~MeV, the values of
 $\beta$ = 82.0$^\circ\pm$0.5$^\circ$ and $\alpha$ = 2.0$^\circ\pm$0.5$^\circ$ were obtained, whereas
 $\beta$ = 99.4$^\circ\pm$0.5$^\circ$ and $\alpha$ = 2.0$^\circ\pm$0.5$^\circ$ were obtained for a deuteron beam energy of 180~MeV.
 Since the sensitivity to the $T_{21}$ analyzing power in Eq.~(\ref{eq2}) goes as $\cos\beta \sin\beta$, as $\beta$ approaches $90^{\circ}$,
 the uncertainty in $T_{21}$ increases. As a consequence, the large systematic errors in
  $T_{21}$ are associated with the value of $\beta$ in these experiments. In addition, variations in the beam
 polarization angles during the experiment have been monitored by dividing the data into several bins in time, and were found to be
 small. Yet these non-statistical fluctuations were taken into account in our $\chi^{2}$ fitting procedure. The total systematic
 error is the quadratic sum of these partial systematic errors which are given in Tabs.~\ref{130table} and \ref{180table}
 with superscript $sys$ and are plotted as gray bands on top of the figures. 

\section {Comparison between data and theory}
The measured vector and tensor analyzing powers are shown in Figs.~\ref{130results} and \ref{180results} by filled circles
 and compared to rigorous Faddeev calculations, represented by bands and lines. The dark gray bands correspond to calculations
 including only two-nucleon potentials, namely Argonne V18 (AV18)~{\cite{AV18}}, Charge-Dependent Bonn (CDB)~{\cite{cdbonn}}, Nijmegen
 I (NIJM~I), and Nijmegen II (NIJM~II)~{\cite{nim93}}. The width of the bands indicates the variation in predictions for the three-body
 systems when using different two-nucleon potentials. The light gray bands represent calculations including an additional TM'
 three-nucleon force as well. The solid lines correspond to the results of a Faddeev calculation using the AV18 two-nucleon potential
 combined with the Urbana-IX three-nucleon potential~{\cite{urbana}}. The dotted lines result from a calculation using the coupled channel
 potential  CDB+$\Delta$~\cite{hannover}. It should be noted that the calculations presented in this paper do not include the Coulomb
 interaction. The Hannover-Lisbon theory group, however, recently provided predictions including the Coulomb
 interaction~\cite{deltacolumb1,deltacolumb2}. For the polarization observables and energies presented here, its effect is found
 to be negligible, and therefore, has been left out to allow a fair comparison with the  other predictions that do not include the
 Coulomb interaction. Also measurements from Ref.~{\cite{witala,garcon} at energies close to our energy are shown by open triangles and
 squares for every energy. They clearly show that the precision of the new measurements has been improved significantly.

A comparison between theory and experiment at $E_d$ =  130 MeV shows that the Faddeev calculations based on a pure two-nucleon
 potential give a reasonable agreement with the measured polarization observables. At this energy,  calculations with the inclusion
 of the well-established TM' three-nucleon force fail to describe the data, in particular for $T_{22}$.
 Calculations using the Urbana~IX three-nucleon potential show, with the exception of $T_{20}$, significant discrepancies with the data
 as well. On the other hand, calculations using the coupled-channel potential, CDB+$\Delta$, demonstrate that the polarization observables
 in the three-nucleon system can be described reasonably well by incorporating consistently an intermediate $\Delta$ resonance which
 seems to have a small effect for all observables except for $T_{21}$. Note that the discrepancies for $T_{21}$ can be partly ascribed 
 to the large systematic uncertainties in this polarization observable.

 At $E_d$ = 180~MeV the behavior of the Faddeev calculations with respect to the data changes drastically. The Faddeev calculations,
 incorporating two-nucleon forces only, show large deviations from the data for all observables. The inclusion of a TM' three-nucleon
 potential remedies these deficiencies significantly for $iT_{11}$ and $T_{20}$. However, the predictions for the tensor analyzing powers,
 $T_{22}$ and $T_{21}$, fail to describe the magnitude and the shape of the measured angular distributions. Furthermore, the model
 including the $\Delta$ resonance fails to describe a large part of these data as well, even though the same model gave a good
 description at 130~MeV.

From Figs.~\ref{130results} and \ref{180results}, we observe that the discrepancies between data and theory depend on the incident
 beam energy. As a further check on this, the beam-energy dependence of our data has been compared with the presently available world
 data base at intermediate energies. Results are shown in Fig.~\ref{compare}. All data points correspond to a fixed angle of $\theta_{c.m.}$ = 100$\pm$0.5$^\circ$ in the center-of-mass. Our data (filled circles) are consistent with the data from Ref.~{\cite{witala}}
 (open triangles). In addition, the precision measurements of Sekiguchi et al.~{\cite{kimiko}} at
 $E_d^{lab}$ = 140 and 200~MeV are plotted as open diamonds. Results of the Faddeev calculations for CDB potential without 3NF and with TM'
 and $\Delta$ are drawn as dashed lines, dash-dotted and solid lines, respectively. 

The large deviations between our data and the rigorous predictions from Faddeev calculations based on modern two-nucleon potentials
 show that the observables shown in this  paper are sensitive to 3NF effects. The role of the $\Delta$-resonance as a degree of freedom
 for 3NF has proved to be the dominant ingredient to describe well the vector analyzing powers in $\vec p+d$ elastic  scattering
 at intermediate energies from 108 MeV to 190 MeV/nucleon~{\cite{ermisch05}}. This paper, however, presents large  discrepancies
 between data and a self-consistent coupled-channel model including a dynamic $\Delta$ resonance at an  incident energy of 90 MeV/nucleon
 for the vector and tensor analyzing powers of the deuteron. For these observables  and energies, the role of the $\Delta$ as a 3NF is
 predicted to be small, whereas the data show that large 3NF effects  are present.  A description of 3NF effects using the
 phenomenological two-pion exchange approach such as the TM' 3NF, which is added to a 2NF, cannot remedy the observed
 discrepancies either. In addition to the models presented in this paper, other approaches are becoming available in the literature.
  One of these, namely chiral-perturbation theory, is based on the symmetries of Quantum Chromodynamics. Within this approach, nuclear
 forces are generated systematically~{\cite{bedaq,eppelbaum}}. Once higher orders are included in the calculations based on this theory
 for intermediate energies, one can make a reasonable comparison with the present data. 
\begin{figure*}[!h]
\centering
\resizebox{1\textwidth}{!}{\includegraphics[angle = 0,width = 1\textwidth]{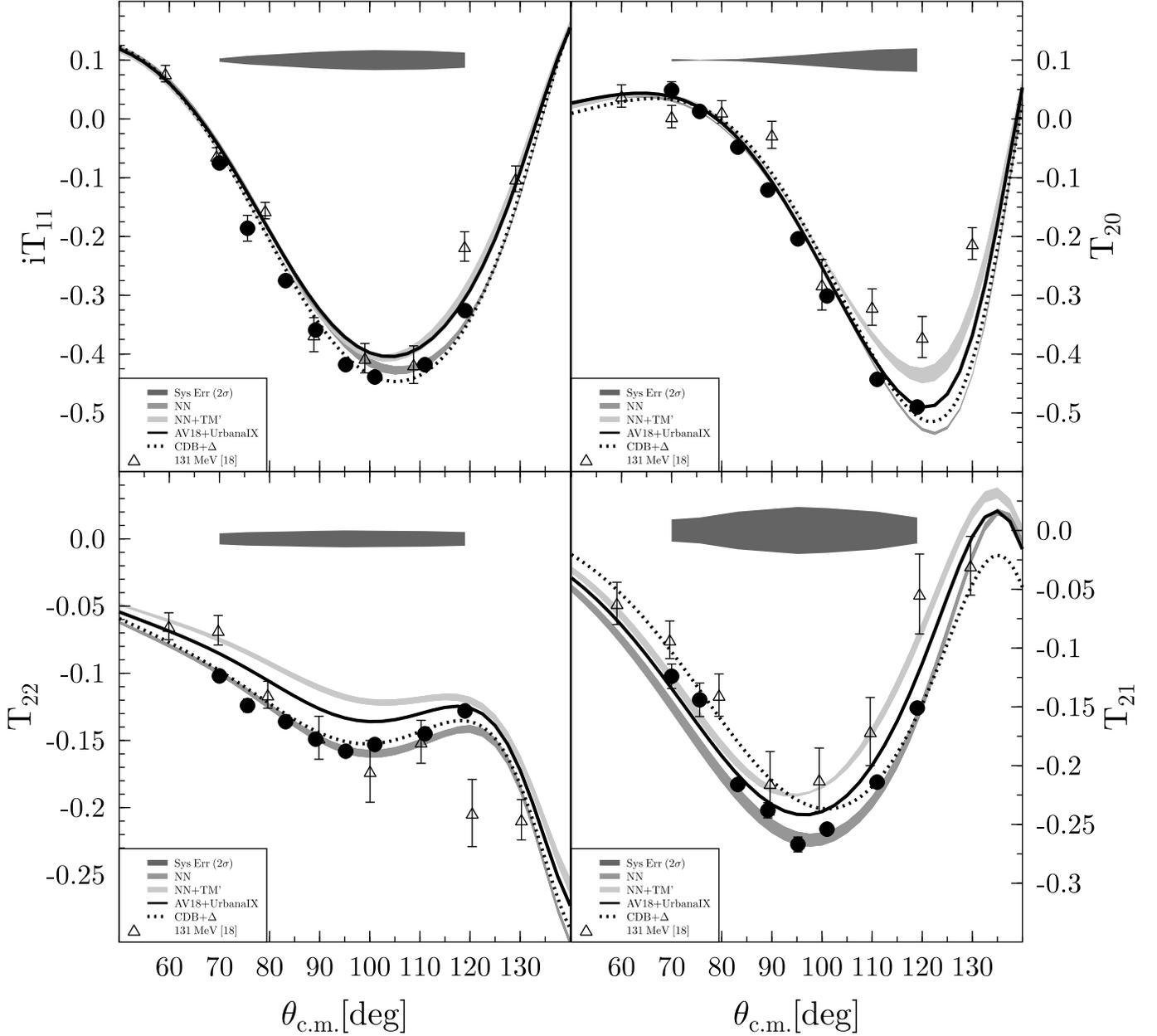}}
\vspace{-5mm}
\caption{Vector and tensor analyzing powers of the elastic $\vec d+p$ scattering at $E^{lab}_{d} = 130$ MeV as a function of
 $\theta_{c.m.}^{\circ}$. Error bars are statistical plus a point-to-point uncertainty added in quadrature. In each panel, filled circles
 are data from the present experiment and open triangles are data from Ref.~{\cite{witala}} at $E^{lab}_{d} = 131$ MeV. The dark gray bands
 at the top of the panels represent the systematical uncertainty ($2\sigma$) for every data point. The other dark gray bands correspond to
 calculations including only two-nucleon potentials. The light gray bands represent calculations including an  additional Tucson-Melbourne
 TM' three-nucleon force as well. The solid lines correspond to results of a Faddeev calculation using the AV18 two-nucleon potential
 combined with the Urbana-IX (UIX) three-nucleon potential. The dotted line represents the results of a coupled-channel potential
 CDB+$\Delta$ calculations.}
\label{130results}
\end{figure*}

\begin{figure*}[!h]
\centering
\resizebox{1\textwidth}{!}{\includegraphics[angle = 0,width = 1\textwidth]{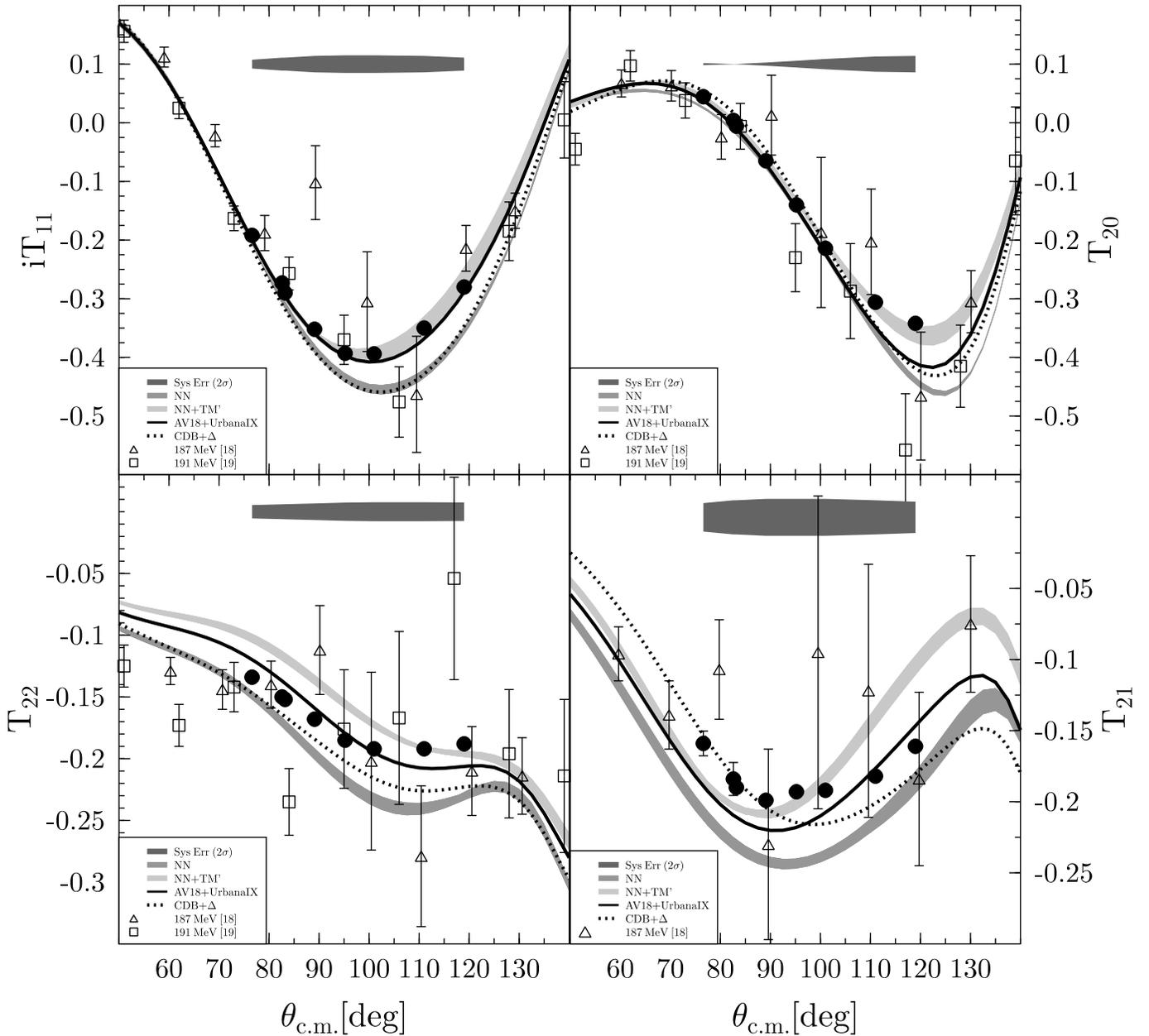}}
\vspace{-5mm}
\caption{Vector and tensor analyzing powers in the elastic $\vec d+p$ scattering at an incident deuteron beam energy of $E^{lab}_{d} = 180$
 MeV. The open triangles are data from Ref.~{\cite{witala}} and the open squares are data from Ref.~{\cite{garcon}}. For a description of the bands and lines, refer to Fig.~\ref{130results}.}
\label{180results}
\end{figure*}

\begin{table*}[!h]
\caption{ Analyzing powers for elastic $\vec d+p$ scattering at $E^{lab}_{d} = 130$ MeV for different scattering angles in the center of mass
 frame. Statistical errors are quoted with the superscript $^{st.}$ and systematical errors with the superscript $^{sys.}$. Statistical errors
 are the output of a $\chi^{2}$ fitting procedure plus an additional point-to-point uncertainty added quadratically. The uncertainty in 
 $\theta_{c.m.}$ is $\pm$0.5$^{\circ}$. Note that all the analyzing power values and errors have been multiplied by a factor 1000.}
\label{130table}
\begin{tabular}{ccccccccccccc}
$\theta_{c.m.}[^{\circ}]$&$iT_{11}$&$\Delta iT_{11}^{st.}$&$\Delta iT_{11}^{sys.}$&$T_{22}$&$\Delta T_{22}^{st.}$&$\Delta T_{22}^{sys.}$&$T_{20}$&$\Delta T_{20}^{st.}$&$\Delta T_{20}^{sys.}$&$T_{21}$&$\Delta T_{21}^{st.}$&$\Delta T_{21}^{sys.}$\\\hline
70   &  $-$75  &   6 &   3  &   $-$102 &   3 &   4 &      49   &  14 &   2  &   $-$124 &  10 &   9  \\      
76   &  $-$186 &  22 &   7  &   $-$124 &   5 &   5 &      13   &   4 &   1  &   $-$144 &  14 &   11  \\      
83   &  $-$275 &   4 &   11 &   $-$136 &   2 &   6 &   $-$48   &   4 &   2  &   $-$216 &   4 &   16  \\      
89   &  $-$359 &   3 &   14 &   $-$149 &   1 &   6 &   $-$121  &   5 &   5  &   $-$238 &   6 &   18  \\      
95   &  $-$418 &   4 &   16 &   $-$158 &   4 &   6 &   $-$204  &   6 &   8  &   $-$267 &   6 &   20  \\      
101  &  $-$439 &   3 &   17 &   $-$153 &   1 &   6 &   $-$301  &   5 &   12 &   $-$254 &   4 &   19  \\      
111  &  $-$418 &   3 &   16 &   $-$145 &   1 &   6 &   $-$443  &   5 &   18 &   $-$214 &   4 &   16  \\      
119  &  $-$326 &   3 &   13 &   $-$128 &   2 &   5 &   $-$490  &   6 &   20 &   $-$151 &   3 &   11  \\\hline
\end{tabular}
\end{table*}

\begin{table*}[!h]
\caption{ Analyzing powers for elastic $\vec d+p$ scattering at $E^{lab}_{d} = 180$ MeV. See Tab.\ref{130table} for more details.}
\label{180table}
\begin{tabular}{ccccccccccccc}
$\theta_{c.m.}[^{\circ}]$&$iT_{11}$&$\Delta iT_{11}^{st.}$&$\Delta iT_{11}^{sys.}$&$T_{22}$&$\Delta T_{22}^{st.}$&$\Delta T_{22}^{sys.}$&$T_{20}$&$\Delta T_{20}^{st.}$&$\Delta T_{20}^{sys.}$&$T_{21}$&$\Delta T_{21}^{st.}$&$\Delta T_{21}^{sys.}$\\\hline
77  &  $-$192 &  7 &  8  &  $-$134 &  3 &  5 &  45     & 7  &  2  &  $-$159 &  9  &  10  \\     
83  &  $-$273 &  9 &  11 &  $-$150 &  4 &  6 &  4      & 8  &  1  &  $-$184 &  12 &  12 \\      
84  &  $-$290 &  2 &  11 &  $-$152 &  1 &  6 &  $-$6   &  1 &  1  &  $-$190 &  3  &  12 \\      
89  &  $-$352 &  3 &  14 &  $-$168 &  1 &  7 &  $-$65  &  1 &  3  &  $-$199 &  3  &  13 \\      
95  &  $-$393 &  3 &  15 &  $-$185 &  1 &  7 &  $-$140 &  2 &  6  &  $-$193 &  3  &  13 \\      
101 &  $-$394 &  3 &  15 &  $-$192 &  1 &  8 &  $-$214 &  3 &  9  &  $-$192 &  3  &  13 \\      
111 &  $-$350 &  3 &  14 &  $-$192 &  2 &  8 &  $-$306 &  3 &  13 &  $-$182 &  7  &  12 \\      
119 &  $-$280 &  3 &  11 &  $-$188 &  2 &  8 &  $-$342 &  4 &  14 &  $-$161 &  4  &  11 \\\hline
\end{tabular}
\end{table*}

\begin{figure*}[!h]
 \centering
\resizebox{1\textwidth}{!}{\includegraphics[angle = 0,width = 1\textwidth]{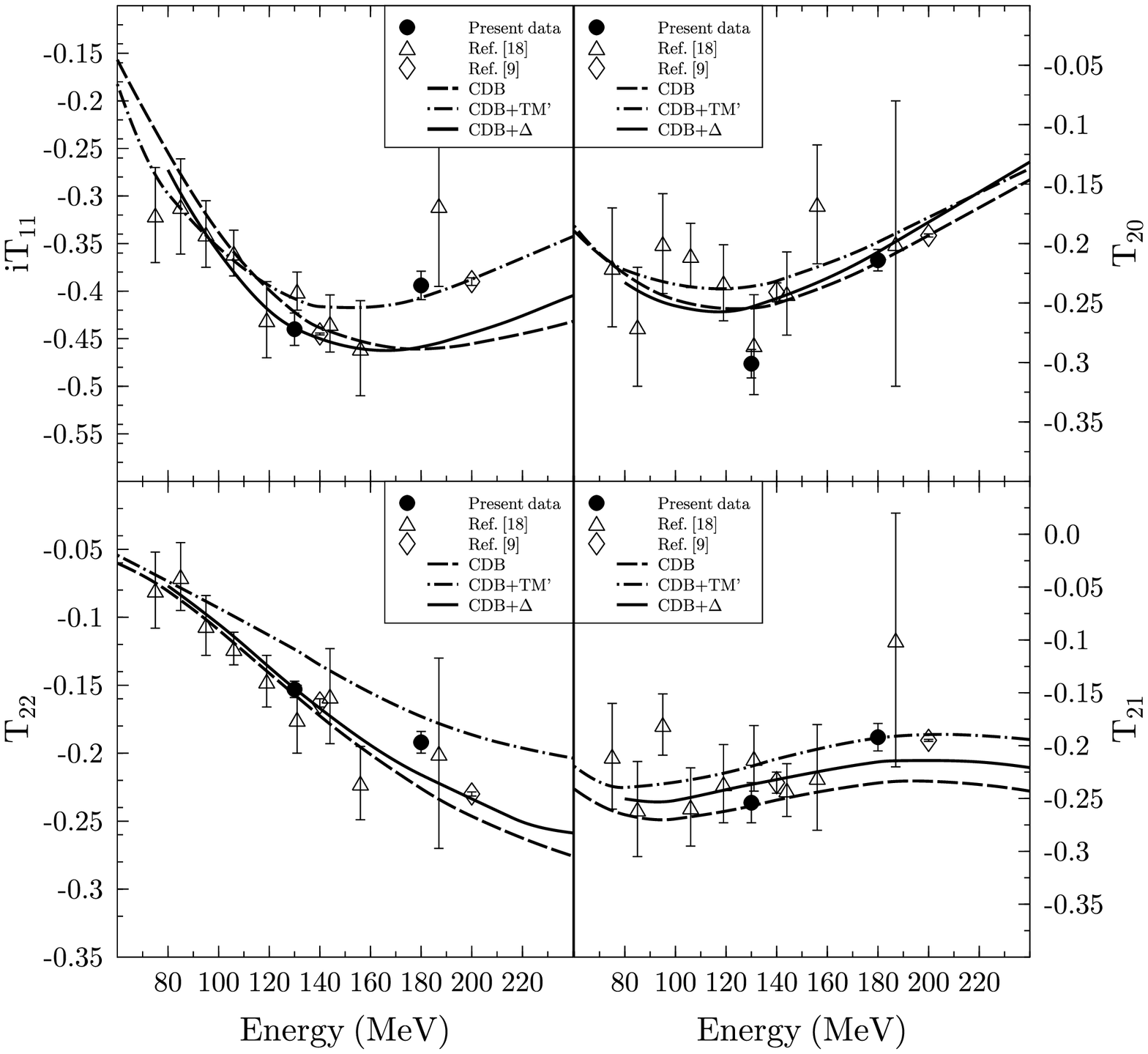}}
\vspace{-5mm}
\caption{A comparison of our measurements of vector and tensor analyzing powers in deuteron-proton elastic scattering with the world
 database at intermediate energies as a function of the incident beam energy. All data points correspond to a center-of-mass angle of
 $\theta_{c.m.}$ = 100$^\circ$. Our data are presented as filled circles, together with that of Ref.~{\cite{kimiko}} (open diamond) and
 of Ref.~{\cite{witala}} (open triangles). The predictions for CDB, CDB+TM', and CDB+$\Delta$  are  represented by dashed, dash-dotted and
 solid lines, respectively.}
\label{compare}
\end{figure*}

\section{Summary and conclusion}
The precision measurement of vector and tensor analyzing powers in elastic $\vec d+p$ scattering has been performed at intermediate
 energies of 130 and 180 MeV in the center of mass angle range of $70^\circ-120^\circ$. Two different setups were used to measure the deuteron
 beam polarization via $^{12}{\rm C}(\vec d,\alpha)^{10}{\rm B}(2^{+})$ reaction and simultaneously the reaction asymmetries via elastic
 $\vec d+p$ reaction. Predictions of various 3NF models have been compared with the data in order to look for evidences of 3NF effects in this energy range.
 At 130 MeV, the agreement between data and calculations including only two-nucleon potentials is better for all observables. Also CDB+$\Delta$
 describes the data better than calculations using other 3NFs. At 180 MeV, the discrepancy between data and calculations is rather large.
 Calculations including TM' agree better for two analyzing powers of $iT_{11}$ and $T_{20}$ but the trend of the data for the other two analyzing powers, $T_{22}$ and $T_{21}$, matches none of the calculations. Our new data points are compared with the existing data base of
 deuteron analyzing powers in the energy range of 75-200 MeV. The energy dependence of all observables confirms that 3NF effects certainly
 exist but their exact nature should be investigated further.

\section{Acknowledgment}
We would like to thank the RIKEN accelerator group for doing an excellent job during this measurement and appreciate RIKEN heavy-ion group for
 their hospitality and help during the experiment. Special thanks go to R.G.E. Timmermans, E. Stephan, St. Kistryn, A. Biegun, U. van Kolck, E. J. Stephenson, and C. Bailey for their valuable discussions and suggestions. This work was performed as part of the research program of the
 ``Stichting voor Fundamenteel Onderzoek der Materie'' (FOM) and was supported by University of Groningen (RUG).

\end{document}